\documentclass[aps,prl,twocolumn,amsmath,amssymb,floatfix,superscriptaddress,showpacs]{revtex4-1}
\usepackage{graphicx}
\usepackage[colorlinks=true,linkcolor=blue,filecolor=blue,urlcolor=blue]{hyperref}

\newcommand{\calI}{\mathcal{I}}
\newcommand{\calIp}{{\mathcal{I}'}}

\newcommand{\calR}{\mathcal{R}}

\newcommand{\calE}{\mathcal{E}}
\newcommand{\calH}{\mathcal{H}}
\newcommand{\ra}{\rightarrow}
\newcommand{\geff}{g_\textrm{eff}}

\DeclareMathOperator{\Tr}{Tr}

\usepackage{amsmath}

\begin{document}

\title{Distinct Trivial Phases Protected by a Point-Group Symmetry \\
in Quantum Spin Chains}

\author{Yohei Fuji}
\affiliation{Institute for Solid State Physics, University of Tokyo, Kashiwa 277-8581, Japan}

\author{Frank Pollmann}
\affiliation{Max-Planck-Institut f\"ur Physik komplexer Systeme, D-01187 Dresden, Germany}

\author{Masaki Oshikawa}
\affiliation{Institute for Solid State Physics, University of Tokyo, Kashiwa 277-8581, Japan}

\date{\today}

\begin{abstract}
The ground state of the $S=1$ antiferromagnetic Heisenberg chain belongs to the Haldane phase -- a well known example of symmetry-protected
topological phase.
A staggered field applied to the $S=1$ antiferromagnetic chain breaks all the symmetries that protect the Haldane phase as a topological phase, reducing it to a trivial phase.
That is, the Haldane phase is then connected adiabatically to an antiferromagnetic product state.
Nevertheless, as long as the symmetry under site-centered inversion combined with a spin rotation is preserved, the phase is still distinct from another trivial phase.
We demonstrate the existence of such distinct symmetry-protected \emph{trivial} phases using a field-theoretical approach and numerical calculations.
Furthermore, a general proof and a non-local order parameter are given in terms of an matrix-product state formulation.
\end{abstract}


\maketitle


\textit{Introduction.---}While symmetry broken phases can be completely classified using the Landau theory, there  still exists no exhaustive understanding of topological quantum phases. 
Topological quantum phases are gapped phases of matter that are distinct from trivially disordered states but cannot be characterized by any local order parameter. 
Over the past few years, new theoretical frameworks have been developed to understand and classify many different topological phases. 
For example, topological phases of noninteracting fermions are now completely classified using K-theory \cite{Kitaev09,Wen12}.
More generally, two gapped ground states belong to the same phase if and only if they can adiabatically connected with respect to local Hamiltonians~\cite{Verstraete05,XChen10,Schuch11}.
Even when different states are connected by a general adiabatic process, it is possible that they can no longer be adiabatically connected if we impose symmetries on the Hamiltonian.
These are either states with spontaneous symmetry breaking or belong to the class of symmetry-protected topological (SPT) phase \cite{ZCGu09,Pollmann10,XChen10,XChen11a,XChen11b,Schuch11,XChen13}.
Examples of SPT phases include  topological insulators \cite{Kane05a}, which are protected by  time reversal symmetry, and the Haldane phase \cite{Haldane83a,*Haldane83b,Affleck87a,*Affleck88} in one dimension, which is protected by either one of the time reversal, bond-centered inversion, or the dihedral group of the spin rotations \cite{Pollmann10}.
In these examples, SPT phases are nontrivial in the sense that they cannot be adiabatically connected to a trivial product state, once an appropriate symmetry is imposed.
They also support gapless edge states and/or nontrivial degenerate structures in the entanglement spectrum.

While the notion of the SPT phases is now established and widely recognized, in this work we demonstrate that site-centered inversion symmetry allows to distinguish different trivial one-dimensional phases.
That is, there are multiple ``symmetry-protected \emph{trivial}'' (SPt) phases, i.e., symmetric phases connected adiabatically to product states, which are still distinct in the presence of the imposed symmetry.
We note that, the word ``symmetry-protected trivial phase'' is sometimes used in place of the standard terminology of SPT (symmetry-protected topological) phase, because the entanglement in such a phase is short-ranged and is removable in an adiabatic process if the symmetry is disregarded.
In contrast, in what is called an SPt phase in this work, the entanglement can be completely removed  adiabatically, even in the presence of the
imposed symmetry, to reduce the state to a product state.
However, it is still distinct from another trivial phase.
The SPt phases introduced here represent a new class of 1D quantum phases that are protected by a point-group symmetry but not captured by the cohomology classification.
While each of these phases is trivial by itself, the quantum phase transition between them is experimentally detectable (e.g., by a divergence of some susceptibilities).
Moreover, we derive non-local order parameters that could be used to characterize each phase.
While the concept of SPt phases is very general, we illustrate it for clarity with a simple model of spin-1 chain in the following.


\textit{The model.---}In order to make the discussion concrete, let us first consider the following simple model of $S=1$ chain:
\begin{align}
\calH & = \sum_i \left[
\vec{S}_i \cdot \vec{S}_{i+1} +
D_z (S^z_i)^2 - h_z (-1)^i S^z_i \right] . 
\label{eq.Ham}
\end{align}
The first term is the standard Heisenberg model with antiferromagnetic exchange interactions which stabilize the celebrated Haldane gap \cite{Haldane83a,*Haldane83b}.
The  $D_z$ term is the uniaxial single-ion anisotropy, which is commonly present in magnetic ions with $S=1$ such as Ni$^{2+}$.
The model with  uniaxial anisotropy $D_z$ and $h_z =0$ is well understood \cite{denNijs89,Tasaki91,WChen03}. 
For small $D_z \ge 0$, the system is in the Haldane phase and undergoes a quantum phase transition into the ``large-$D$'' phase at $D_z\approx1$. 
Both phases are gapped and have the full symmetry of the Hamiltonian.
The Haldane phase is a well-known example of SPT phases \cite{Pollmann10,Pollmann12a} as discussed in the introduction. 
The large-$D$ phase is a trivial phase which is adiabatically connected to the product state $\left| D \right\rangle = \left| \cdots 0000 \cdots \right\rangle$, where $|0\rangle$ represents the local spin state with $S^z=0$. 
This state is  the exact ground state of the Hamiltonian~\eqref{eq.Ham} in the limit $D_z\rightarrow\infty$.
The $h_z$ term represents a staggered field, which occurs in many quasi-one-dimensional materials, including Haldane gap systems, with an alternating crystal structure under an applied (uniform) field.
For simplicity, we only include the staggered field term without the uniform one.

In the limit $h_z \to \infty$, the spins are fully polarized along the staggered field, and the ground state is reduced to another trivial product state $\left| N \right\rangle = \left| \cdots +-+- \cdots \right\rangle$, where $+$ and $-$ represent the local spin states with $S^z=+1$ and $S^z=-1$, respectively.
It was recognized earlier that there is no phase transition for $0 < h_z < \infty$ (for $D_z =0$) \cite{Tsukano98a}.
That is, the Haldane phase is adiabatically connected to the N\'eel state $|N\rangle$ with imposed antiferromagnetic (AF) order.
In the SPT framework, this is naturally understood;
since the staggered field breaks all the symmetries that protect the Haldane phase as an SPT phase, it reduces the Haldane phase to a trivial phase which also includes the N\'{e}el state $|N\rangle$.

Now let us discuss the model with both $D_z$ and $h_z$.
In fact, this model has been studied in Ref.~\cite{Tsukano98b} by a field theory and numerical methods where a quantum phase transition between the large-$D$ phase and the imposed AF phase was found.
This is rather surprising, since both phases are trivial and are adiabatically connected to product states $|D\rangle$ and $|N\rangle$ that have the full symmetry of the Hamiltonian.
It is perhaps even more surprising in the light of the recent concept of the SPT phases, where the existing classification scheme \cite{XChen11a,XChen11b,Schuch11} would not distinguish them.
While the nature of the phase transition were studied in Ref.~\cite{Tsukano98b}, why (and when) these two trivial phases are distinguished was not completely clarified.
In the remainder of this paper, we demonstrate that this is an example of distinct SPt phases and identify the symmetry which protects them.


\textit{Bosonization.---}
The standard bosonization procedure of $S=1$ chains starts from two coupled $S=1/2$ chains, and the low-energy effective field theory for
$\calH$ is given by the Hamiltonian \cite{Schulz86,Berg08,Berg11}
\begin{align}
H_\textrm{eff} &= \frac{v}{2\pi} \int dx \; \left[ K (\partial_x
\theta)^2 +\frac{1}{K} (\partial_x \phi)^2 \right] \notag \\ & + \geff
\int dx \cos{(2\phi)}, \label{eq:EffHam}
\end{align}
where $\phi$ and $\theta$ are dual field of each other, satisfying $[\phi(x), \theta(x')] = i(\pi/2)[\textrm{sgn} (x-x')+1]$.
The Hamiltonian~\eqref{eq:EffHam} represents the so-called sine-Gordon field theory, which is ubiquitous in many problems in $1+1$ dimensions.
Its properties essentially depend on the coupling constant $K$. 
When $K>2$, the coupling $\geff$ is irrelevant under the renormalization group (RG), and the system is renormalized in the low-energy limit into the free boson theory with $\geff=0$, which is nothing but a gapless Tomonaga-Luttinger liquid (TLL).
On the other hand, if $K<2$ and $\geff$ is non-vanishing, the interaction is RG-relevant, and the system acquires an excitation gap.
In the absence of the staggered field, the Haldane and the large-$D$ phases correspond to $\geff>0$ and $\geff<0$ respectively, both with $K<2$.
It is easy to see that, within the effective Hamiltonian~\eqref{eq:EffHam}, the two phases with $\geff>0$ and $\geff<0$ are always separated by the critical point $\geff=0$.
This actually comes from the fact that $\cos(2\phi)$ is the only interaction compatible with the symmetry and the compactification, $\phi \sim \phi +\pi$ and $\theta \sim \theta+2\pi$, up to subleading terms $\cos(2n\phi)$ $(n \geq 2)$. 
In general, the effective theory can also have the $\sin{(2\phi)}$ term, which can be combined with the $\cos{(2\phi)}$ term as $\cos{(2 \phi + \alpha)}$ with a phase shift $\alpha$.
It is clear that, by changing $\alpha$ from $0$ to $\pi$, the two phases with $\geff>0$ and $\geff<0$ are adiabatically connected without closing the gap \cite{Berg08}.
Thus, for the two phases to be distinct, $\sin{(2\phi)}$ has to be forbidden by some symmetry.

\begin{table*}[t]
\caption{Symmetry transformations for the original spins and the bosonic fields.}
\label{table:SymOp}
\begin{ruledtabular}
\begin{tabular}{lccc}
Symmetry operation & Symbol & Transformation for spins & Transformation for fields $(\phi, \theta)$ \\ \hline
Bond-centered inversion & $\calI_b$ & $\vec{S}_i \ra \vec{S}_{1-i}$ & $\phi (x) \ra -\phi (-x)$, $\theta (x) \ra \theta (-x) +\pi$ \\
Site-centered inversion & $\calI_s$ & $\vec{S}_i \ra \vec{S}_{-i}$ & $\phi (x) \ra -\phi (-x) +\pi$, $\theta (x) \ra \theta (-x)$ \\
$\pi$ rotation about $z$ axis & $\calR_z$ & $S^{x,y}_i \ra -S^{x,y}_i$, $S^z_i \ra S^z_i$ & $\phi \ra \phi$, \ $\theta \ra \theta +\pi$ \\
\end{tabular}
\end{ruledtabular}
\end{table*}

In fact, in the framework of bosonization, this is how a symmetry protects the Haldane phase as an SPT phase which is distinct from the trivial large-$D$ phase.
Any of the three symmetries, which are known to protect the Haldane phase, forbids the $\sin{(2\phi)}$ interaction \cite{Fuji14}.
Here, for brevity, among these three symmetries, we only show the representation of the bond-centered inversion $\calI_b$ in terms of the bosonic field $\phi$, $\theta$ in Table~\ref{table:SymOp}.
The action of $\calI_b$, $\phi(x) \to - \phi(-x)$, forbids $\sin{(2 \phi)}$, which leads to the distinction of the two phases with $\geff>0$ and $\geff<0$.
On the other hand, Table~\ref{table:SymOp} shows that the site-centered inversion $\calI_s$ also has the same action on $\phi$, forbidding $\sin{(2\phi)}$.
However, $\calI_s$ by itself is not sufficient to keep the distinction between the two phases;
they are adiabatically connected without gap closing, because of the vertex operators $e^{\pm i \theta}$ allowed in the absence of the U(1) symmetry of spin rotation about $z$ axis.
According to Table~\ref{table:SymOp}, the combined operation $\calI' = \calI_s \times \calR_z$ ($\calR_z$ is the global $\pi$-rotation about $z$ axis), as well as $\calI_b$, forbids $e^{\pm i \theta}$.
They still allow the next leading ones $e^{\pm 2 i \theta}$.
Nevertheless, they just replace the direct transition between the $\geff >0$ and $\geff <0$ phases by an intermediate phase with spontaneous $\mathbb{Z}_2$-symmetry breaking.
Thus $\calI'$ alone should still maintain the distinction between two SPt phases.


\textit{Numerical results.---}The above bosonization analysis suggests that we can introduce microscopic models with less symmetries than Eq.~\eqref{eq.Ham}, but with the symmetry under $\calI'$, to maintain the two distinct phases. 
As an example, we consider the following Hamiltonian:
\begin{align}
\calH' = \calH + \sum_i d_x \left( S^y_i S^z_{i+1} - S^z_i S^y_{i+1} \right)
\label{eq:Spin1Model}
\end{align}
The new term $d_x$ represents the (uniform) Dzyaloshinskii-Moriya (DM) interaction with the DM vector parallel to $x$ axis.
This term breaks not only the U(1) spin-rotational symmetry about $z$ axis, but also both $\calI_s$ and $\calR_z$ as individual symmetries.
However, $\calH'$ with a non-vanishing $d_x$ still preserves the symmetry $\calI'$ under the composite operation.

\begin{figure}
\includegraphics[clip,width=0.45\textwidth]{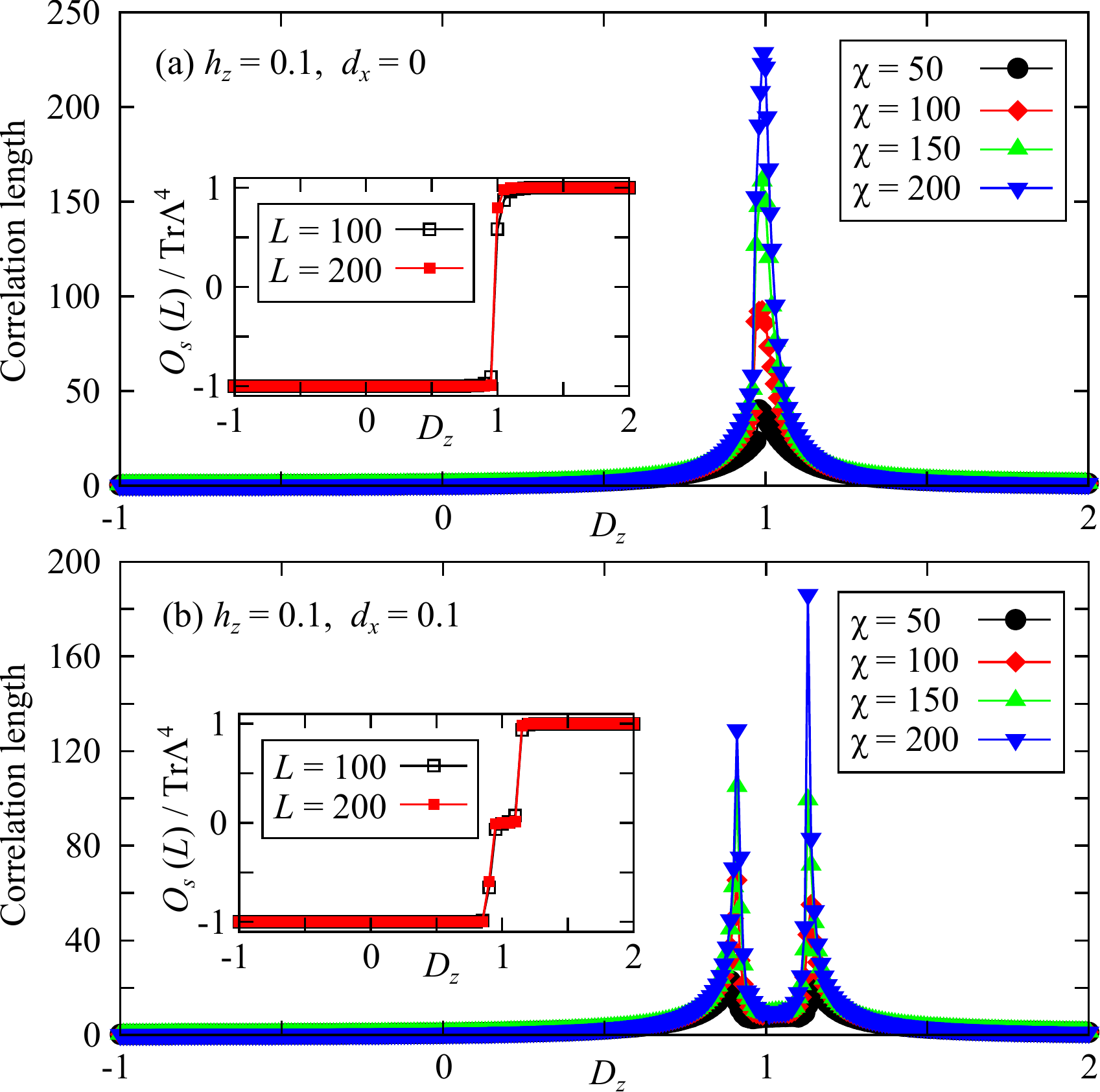}
\caption{(Color online) Correlation lengths calculated for the spin-1 chain \eqref{eq:Spin1Model} are plotted against $D_z$. 
The parameters are varied as (a) $h_z=0.1$, $d_x=0$, and (b) $h_z=d_x=0.1$. 
Each color and symbol denotes the different number of kept states $\chi$ from $50$ to $200$.
Insets show the nonlocal order parameters $\mathcal{O}_s(L)/\Tr \Lambda^4$ for $L=100$ and $200$ (see text).}
\label{fig:CorrLength}
\end{figure}

We numerically study Hamiltonian~\eqref{eq:Spin1Model} using infinite density-matrix renormalization group (iDMRG) \cite{White92,McCulloch08,Kjall13}.
The correlation lengths as functions of $D_z$ are plotted in Fig.~\ref{fig:CorrLength} for different numbers $\chi$ of kept states and parameters of the model.  
A divergent correlation length with increasing $\chi$ indicates a critical point.  
For $h_z=0.1$ and $d_x=0$, we find that the Haldane phase and the N\'eel state ($D_z \rightarrow -\infty $) are adiabatically connected since all of the three symmetries protecting the Haldane phase are broken.  
However, as found in Refs.~\cite{Tsukano98b,XDeng13}, the transition at $D_z \sim 1$ still exists (see Fig.~\ref{fig:CorrLength}~(a)).
This indicates that there is a phase transition between two trivial phases connected to $| N \rangle$ and $| D \rangle$.
To confirm that this transition is protected by $\calI'$ alone, we further introduce $d_x$ in Fig.~\ref{fig:CorrLength}~(b).
A single transition in Fig.~\ref{fig:CorrLength}~(a) is now split into two transitions, but the two phases are still separated by (two) transitions and thus are distinct.
In the intermediate phase, an AF order along $x$ axis occurs and thus $\calI'$ is spontaneously broken. 
Further details about this calculation are shown in \footnotemark[2].

Once an explicit dimerization in introduced, e.g., by adding a term $\delta \sum_i (-1)^i \vec{S}_i \cdot \vec{S}_{i+1}$ with $\delta \neq 0$, $\calI'$ is broken without affecting any other symmetries in $\calH'$, and there is only one trivial phase. 
Numerically, we observe that the correlation length remains finite for all values of $D_z$ when $d_x=0$ \footnotemark[2].
In fact, this can also be shown analytically by considering the limit of $\delta=1$ with $d_x =0$.
Here, the entire chain is decomposed into isolated dimers. 
In particular, at $D_z = h_z =0$, the ground state is simply given by a product of spin singlet states on each dimer.
It can be shown, by solving the two-spin problem explicitly \footnote[2]{See Supplemental Material for details, which includes Refs.~\cite{Calabrese04,ZXLiu12}.}, that this dimerized state is connected adiabatically to both $D_z \to \infty$ and $h_z \to \infty$ limits.
Thus, the two trivial product states $|D\rangle$ and $|N\rangle$ can be adiabatically connected through the dimerized limit, and belong to a single phase, in the presence of $\delta$.
This fact rules out the possibility that the two trivial phases are distinct under the two-site translation invariance and some on-site symmetry, as indicated in Refs.~\cite{XChen11a,XChen11b}. 


\textit{Matrix-product state formulation.---}Matrix-product states (MPS) can represent gapped ground states of local Hamiltonians in one dimension faithfully.
Thus, the classification of gapped phases in one dimension, including the SPt phase proposed in the present work, can be proven rigorously within the MPS formalism.
Let us begin with the general MPS \cite{Vidal03,Orus08}, without assuming any translation invariance:
\begin{eqnarray}
\left| \psi \right\rangle &=& \sum_{\{m_n\}} \cdots \Gamma^{[n-1]}_{m_{n-1}}
\Lambda^{[n-\frac{1}{2}]} \Gamma^{[n]}_{m_n} \Lambda^{[n+\frac{1}{2}]}
\Gamma^{[n+1]}_{m_{n+1}} \cdots \nonumber \\ && \times \left| \cdots
m_{n-1} m_n m_{n+1} \cdots \right\rangle,
\end{eqnarray}
where $\Lambda^{[a]}$ is a $\chi_a \times \chi_a$ positive diagonal matrix, $\Gamma^{[n]}$ is a $\chi_{n-1/2} \times \chi_{n+1/2}$ matrix, and $m_n$ represents the physical degrees of freedom on site $n$.
An MPS representation is not unique for a given state but we can always choose the canonical MPS~\cite{PerezGarcia07} satisfying  $\Tr{\left[\left(\Lambda^{[a]}\right)^2\right]} =1$, and
\begin{equation}
\calE^{[n]}(\mathbb{I}_{\chi_{n+1/2}})  = \mathbb{I}_{\chi_{n-1/2}},
\; \;
\bar{\calE}^{[n]} (\mathbb{I}_{\chi_{n-1/2}}) =
 \mathbb{I}_{\chi_{n+1/2}},
\label{eq.canonical}
\end{equation}
where $\mathbb{I}_{\chi}$ is the $\chi \times \chi$ identity matrix, and $\calE^{[n]}$ and $\bar{\calE}^{[n]}$ are completely positive maps defined by
\begin{equation}
\begin{split}
\calE^{[n]}(X) & \equiv \sum_m \Gamma_m^{[n]} \Lambda^{[n+\frac{1}{2}]} X
\Lambda^{[n+\frac{1}{2}]} \left( \Gamma_m^{[n]} \right)^\dagger, \\
\bar{\calE}^{[n]} (Y) & \equiv \sum_m \left( \Gamma_m^{[n]} \right)^\dagger
\Lambda^{[n-\frac{1}{2}]} Y \Lambda^{[n-\frac{1}{2}]} \Gamma_m^{[n]}.
\end{split}
\end{equation}
By introducing the metric  $|X|^2 \equiv \textrm{Tr}[ X (\Lambda^{[a]})^2 X^\dagger]$ in the vector space of $\chi_a \times \chi_a$ matrices, we can introduce a singular value decomposition of $\calE^{[n]}$ and $\bar{\calE}^{[n]}$.
The canonical condition Eq.~\eqref{eq.canonical} means that the identity matrices are left/right ``eigenvectors'' of $\calE^{[n]}$ and $\bar{\calE}^{[n]}$ belonging to the largest singular value $1$.
In the following we assume that the MPS is pure, that is the largest singular value $1$ is nondegenerate~\footnotemark[2].
In order to consider the symmetry $\calI'$, we define $n \in \mathbb{Z}$ so that $\calI_s$ can be identified with $n \to -n$ with the inversion center at site $n=0$.
Following Refs.~\cite{PerezGarcia08,Pollmann10}, if $\left| \psi \right>$ is invariant under the combined symmetry $\calI'$, it satisfies
\begin{align}
\sum_{m'} u_{mm'} \left( \Gamma_{m'}^{[n]} \right)^T =
e^{i\theta_{\calI'}^{[n]}} \left( U_{\calI'}^{[-n-\frac{1}{2}]}
\right)^\dagger \Gamma_m^{[-n]} U_{\calI'}^{[-n+\frac{1}{2}]},
\label{eq.fe_calIp}
\end{align}
where $u_{mm'}$ is the representation of $\calR_z$ acting on the physical Hilbert space of each site, $\theta_{\calI'}^{[n]}$ is a phase, and $U_{\calI'}^{[a]}$ is a $\chi_a \times \chi_a$ unitary matrix commuting with $\Lambda^{[a]}$.  $\calIp$ also implies that $\Lambda^{[a]} = \Lambda^{[-a]}$ and $\chi_a = \chi_{-a}$.
Using the above relation twice, we obtain
\begin{align}
 \bar{\calE}^{[n]} \left( A^{[n-\frac{1}{2}]} \right)
& = e^{- i (\theta^{[n]}_\calIp + \theta^{[-n]}_\calIp)}
    A^{[n+\frac{1}{2}]},
\label{eq.cmpEA}
\end{align}
where $ A^{[a]}  \equiv \left( U^{[-a]}_\calIp\right)^T \left( U^{[a]}_\calIp\right)^\dagger$.
Since $|A^{[a]}|^2=1$, Eq.~\eqref{eq.cmpEA} implies $A^{[n\pm 1/2]}$ are left/right eigenvectors of $\bar{\calE}^{[n]}$ belonging to the singular value $1$.
The assumption of the pure MPS, namely non-degeneracy of the singular value $1$ of $\bar{\calE}^{[n]}(X)$, implies $A^{[a]} = e^{ i \phi^{[a]}_\calIp} \mathbb{I}_{\chi_a}$, where $\phi^{[a]}_\calIp + \phi^{[-a]}_\calIp = 0\mod{2\pi}$.
Combining these with Eq.~\eqref{eq.cmpEA} and the canonical condition, we obtain $\theta^{[n]}_\calIp + \theta^{[-n]}_\calIp = \phi^{[n+1/2]}_\calIp - \phi^{[n-1/2]}_\calIp \mod{2\pi}$.
In particular, for $n=0$, we find
\begin{align}
 2 (\theta^{[0]}_\calIp - \phi^{[1/2]}_\calIp) = 0 \mod{2\pi}.
\end{align}
As a consequence, $\theta^{[0]}_\calIp-\phi^{[1/2]}_\calIp$ is quantized to either $0$ or $\pi$; it cannot change unless the system undergoes a quantum phase transition. 
This implies that, in the presence of the $\calIp$ symmetry, there are two distinct phases corresponding to $\theta^{[0]}_\calIp-\phi^{[1/2]}_\calIp =0$ and $\pi$.
Let us now consider the limits of the trivial product states $|D\rangle$ and $|N\rangle$.
Here, all the matrices $\Gamma^{[n]}$, $\Lambda^{[n]}$, and $U^{[a]}_\calIp$  are reduced to scalars ($1 \times 1$ matrices) and thus commute with each other.
Then the fundamental relation~\eqref{eq.fe_calIp} for $n=0$ reads $\theta^{[0]}_\calIp-\phi^{[1/2]}_\calIp = 0$ for $|D\rangle$ and $\theta^{[0]}_\calIp-\phi^{[1/2]}_\calIp = \pi$ for $|N\rangle$.
This establishes that, under the $\calI'$ symmetry, the two product states $|D\rangle$ and $|N\rangle$ indeed belong to distinct phases, which are always separated by a quantum phase transition.

As in the case of SPT phases, no local order parameter can distinguish SPt phases. However, using the MPS framework, we can directly derive \emph{non-local} order parameters \cite{Pollmann12b} which are sensitive to the phase factor $\theta^{[0]}_\calIp-\phi^{[1/2]}_\calIp$.
In particular, we can define an operator $\mathcal{I}'_s(2L+1)$ which inverts a block of $2L+1$ consecutive sites.
For $L$ much larger than the correlation length, we find that 
\begin{align}
\mathcal{O}_s(L) = \langle\psi|\mathcal{I}'_s(2L+1)|\psi\rangle \approx \Tr \Lambda^4 e^{i(\theta^{[0]}_\calIp-\phi^{[1/2]}_\calIp)}. 
\end{align}
The insets in Fig.~\ref{fig:CorrLength} show that the different SPt phases are indeed distinguished by $\mathcal{O}_s(L)/\Tr \Lambda^4= \pm 1$ while $\mathcal{O}_s(L)=0$ when $\calI'$ is broken. 


\textit{Conclusion and discussion.---}
We demonstrated that there exists two distinct SPt phases in the presence of the symmetry under the site-centered inversion combined with a spin rotation.
We showed the existence of such phases by field-theoretical arguments based on bosonization and presented a general proof based on the MPS formalism.
While it is known that distinct trivial phases can exist in translation-invariant systems \cite{XChen11a,XChen11b}, it is surprising that only point-group symmetries can stabilize distinct trivial phases in simple 1D systems.
Our finding implies that more studies are needed for complete classification of quantum phases in one dimension under symmetries.
We derived non-local order parameter that could be measured in optical lattice realizations \cite{Endres11}.
Moreover, even without any measurement of the non-local order parameter, the quantum phase transition separating the distinct SPt phases can be detected in standard experimental measurements, such as a divergence of the low-temperature specific heat when the gap is closing.
This, in fact, is more in line with the operational definition of the SPt phases.

The notion of the SPt phases is not restricted to one-dimensional systems.
In fact, what appear as examples of SPt phases in two dimensions were discussed in Ref.~\cite{HYao07,HYao10}.
The lack of universal theoretical description of quantum many-body systems in higher dimensions makes a systematic analysis of SPt phases more difficult than in one dimension.
Nevertheless, it would be certainly an interesting direction for the future.

\paragraph{Acknowledgment---}
YF thanks P. Lecheminant, S. Nishimoto, and K. Totsuka for fruitful discussions. 
YF was supported in part by the Program for Leading Graduate Schools, MEXT, Japan.
This work was supported in part by JSPS Grant-in-Aid for Scientific Research (KAKENHI) No. 25400392 and JSPS Strategic International Networks Program No. R2604 ``TopoNet.''
Numerical calculations were performed on supercomputers at the Institute for Solid State Physics, University of Tokyo.

%


\clearpage
\setcounter{equation}{0}
\setcounter{figure}{0}
\setcounter{table}{0}
\makeatletter

\renewcommand{\theequation}{S\arabic{equation}}
\renewcommand{\thefigure}{S\arabic{figure}}
\renewcommand{\bibnumfmt}[1]{[S#1]}
\renewcommand{\citenumfont}[1]{S#1}

\onecolumngrid
\section*{Details for numerical calculation}

In this supplemental material, we present several numerical details about the spin-1 chain, 
\begin{eqnarray}
H = \sum_i \left[ \left( 1+\delta (-1)^i \right) \vec{S}_i \cdot \vec{S}_{i+1} + D_z (S^z_i)^2 +h_z (-1)^i S^z_i + d_x (S^y_i S^z_{i+1} -S^z_i S^y_{i+1}) \right]. 
\end{eqnarray}

\subsection{Central charge, magnetization, and entanglement spectrum}

\begin{figure*}[b]
\includegraphics[clip,width=0.8\textwidth]{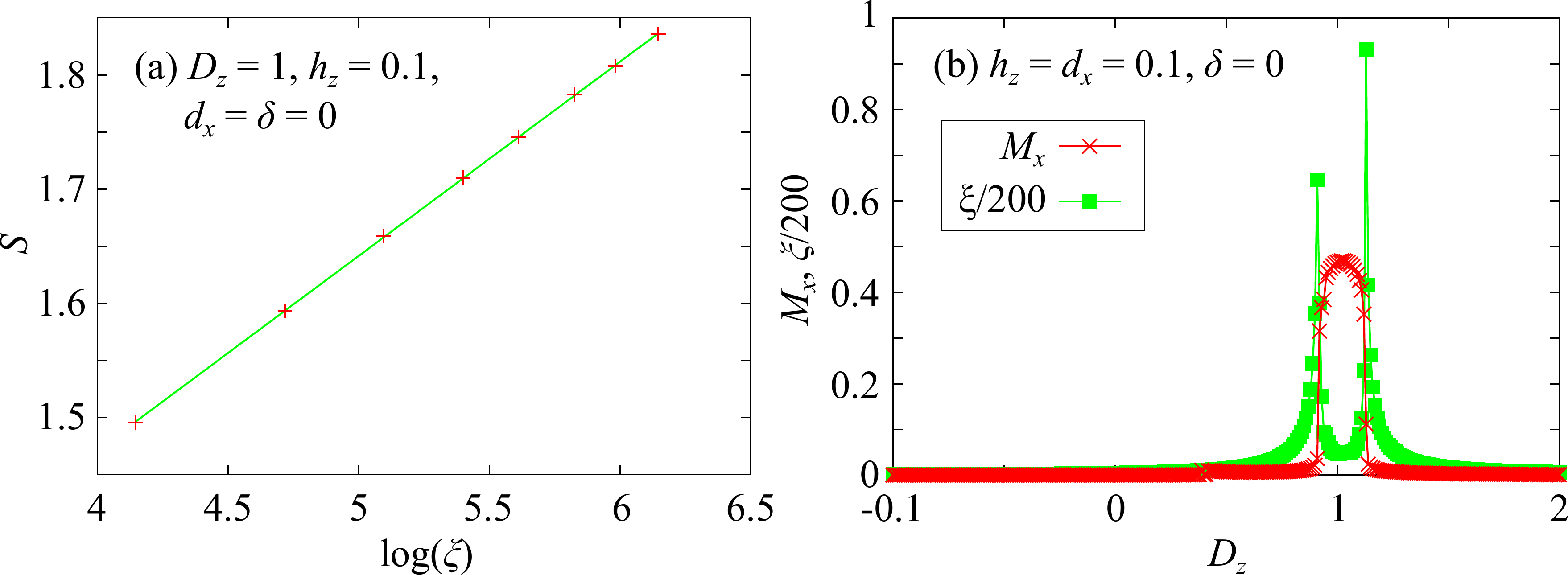}
\caption{(a) von Neumann entanglement entropy as a function of the correlation length $\xi$ for $D_z=1$, $h_z=0.1$, and $d_x=\delta=0$. 
The solid line is a logarithmic fitting function $S=0.170 \log \xi+0.792$. 
(b) Staggered magnetization in $x$ axis as a function of $D_z$ for $\chi=200$, $h_z=d_x=0.1$, and $\delta=0$. 
The correlation length (divided by 200) is again shown for comparison. }
\label{fig:EEandMx}
\end{figure*}

For $h_z=0.1$ and $d_x=\delta=0$, from the divergent behavior of the correlation length in Fig.1(a) of the main text, we expect a Gaussian transition at $D_z \sim 1$. 
To confirm this, we calculate the von Neumann entanglement entropy $S$ for a bipartition of the system into two half chains as a function of the correlation length $\xi$. 
From conformal field theory, the entanglement entropy is known to scale as \cite{Calabrese04S,Kjall13S}
\begin{eqnarray}
S = \frac{c}{6} \log (\xi /a) + c',
\end{eqnarray}
where $c$ is the central charge, $a$ is the lattice spacing (we set $a=1$), and $c'$ is a nonuniversal constant. 
As shown in Fig.~\ref{fig:EEandMx} (a), the entanglement entropy $S$ at $D_z=1.00$ is well fitted by a linear function of $\log (\xi)$, and the central charge is estimated as $c \approx 1.02$. 
This is close to the expected value $c=1$ at the Gaussian transition. 

For $h_z=d_x=0.1$ and $\delta=0$, a Gaussian transition at $d_x=0$ splits into two Ising transitions, and we have an intermediate phase between these transitions. 
In fact, between the two peaks in Fig.1(b) of the main text, the $\mathbb{Z}_2$ spin reversal symmetry in $x$ axis ($S^x_i \rightarrow -S^x_i$) is spontaneously broken. 
Then the staggered magnetizations along $x$ axis, $M_x$, take finite expectation values in the intermediate phase, as shown in Fig.~\ref{fig:EEandMx} (b). 
Such an intermediate N\'eel phase between two distinct gapped symmetric phases is also observed in a two-leg spin-$1/2$ ladder \cite{ZXLiu12S} when the $U(1)$ symmetry is explicitly broken. 

\begin{figure*}
\includegraphics[clip,width=0.9\textwidth]{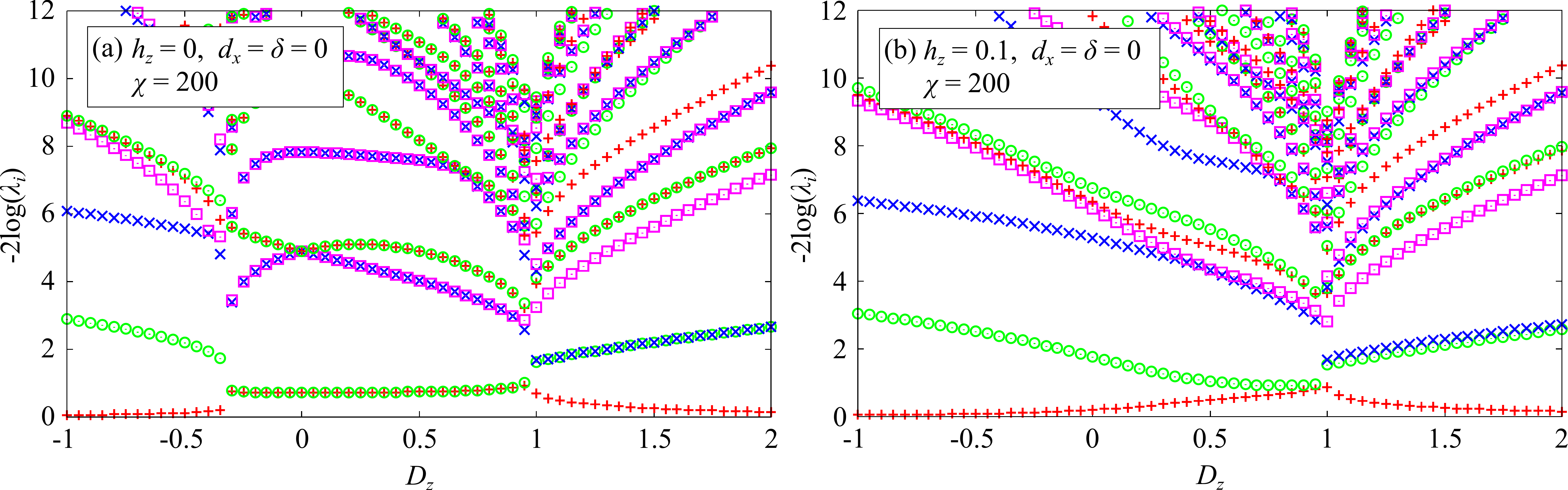}
\caption{Low-lying entanglement spectra as functions of $D_z$ for (a) $h_z=d_x=\delta=0$ and (b) $h_z=0.1$, $d_x=\delta=0$. 
Both data are obtained with $\chi=200$. }
\label{fig:EntSpec}
\end{figure*}

To see a clear signature of the topologically trivial phases, we check the degeneracy in the entanglement spectra. 
For $h_z=d_x=\delta=0$, the Haldane phase is protected by time reversal, bond-centered inversion, and dihedral group of the spin rotations, as observed in Ref.~\cite{Pollmann10S}. 
From Fig.~\ref{fig:EntSpec} (a), the whole entanglement spectrum is two-fold degenerate in a region $-0.3 \lesssim D_z \lesssim 1$. 
On the other hand, as shown in Fig.~\ref{fig:EntSpec} (b), once we introduce a finite staggered magnetic field $h_z$, this two-fold degeneracy is lifted, and the Haldane phase is merged with the imposed AF ordered phase which is topologically trivial. 

\subsection{Adiabatic connection between \texorpdfstring{$\left| D \right>$ and $\left| N \right>$}{|D> and |N>}}

\begin{figure}
\begin{center}
\includegraphics[clip,width=0.5\textwidth]{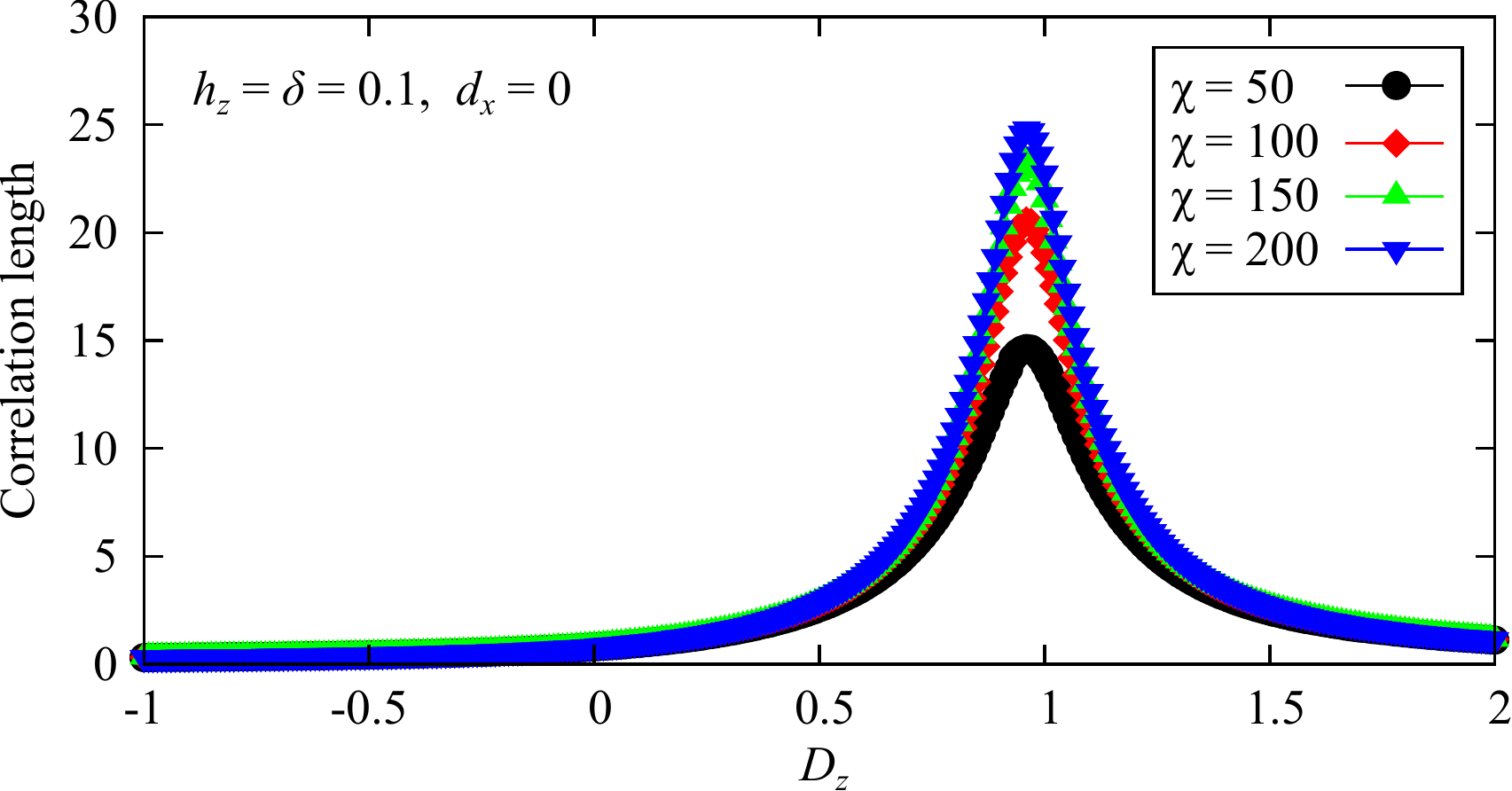}
\caption{Correlation length as a function of $D_z$ for the spin-1 chain with $h_z=\delta=0.1$ and $d_x=0$. 
Each color and symbol denotes the different number of $\chi$ varied from $50$ to $200$.}
\label{fig:CL_Dimer}
\end{center}
\end{figure}

A nonzero dimerization $\delta$ breaks the site-centered inversion symmetry but still preserves the two-site translational invariance. 
In the main text, we discuss an adiabatic continuity between the two trivial states $| D \rangle$ and $| N \rangle$ in the absence of the combined symmetry, $\mathcal{I}' = \mathcal{I}_s \times \mathcal{R}_z$. 
We first show the correlation length as a function of $D_z$ for $h_z=\delta=0.1$ and $d_x=0$ in Fig.~\ref{fig:CL_Dimer}. 
Compared with Fig.1(a) in the main text, the correlation length around $D_z \simeq 1$ becomes shorter and exhibits a saturating behavior by increasing $\chi$. 
This indicates the absence of the phase transition due to the dimerization which breaks the combined symmetry $\mathcal{I}'$. 
\begin{figure}
\begin{center}
\includegraphics[clip,width=0.7\textwidth]{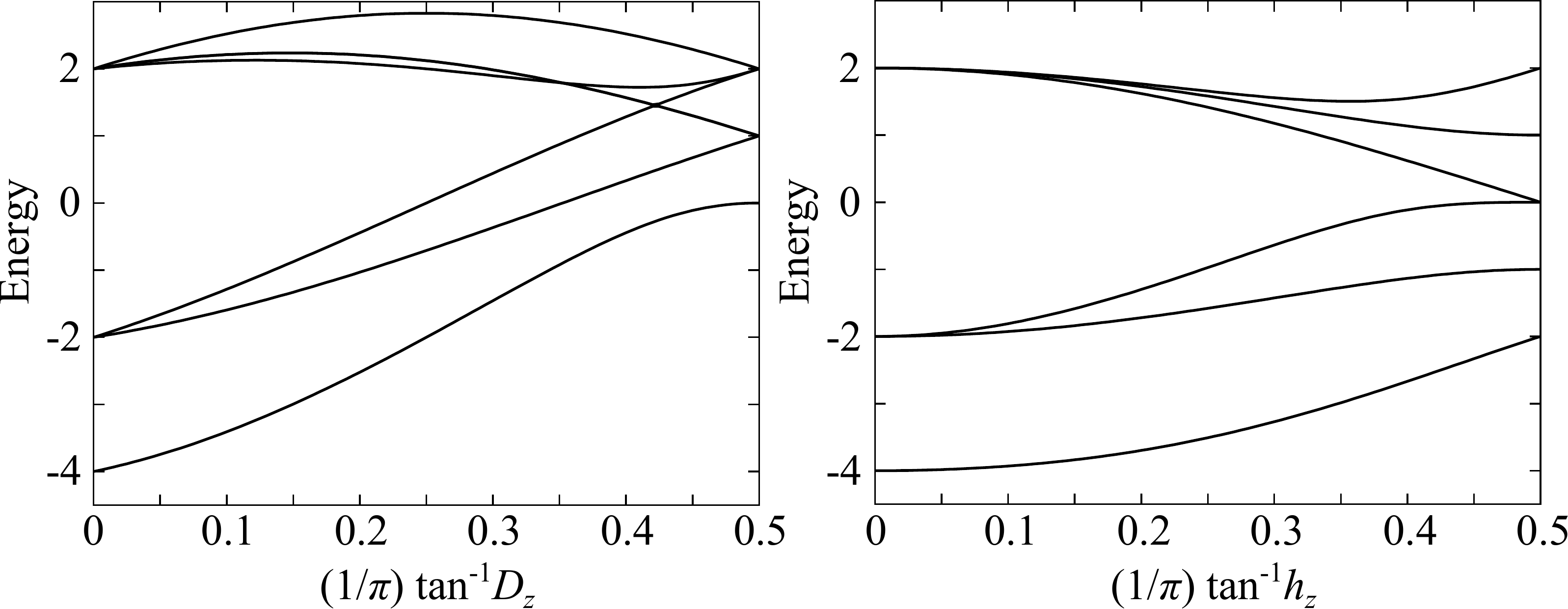}
\caption{Energy spectra of the two-site Hamiltonian \eqref{eq:2body}. 
$D_z$ are varied and $h_z=0$ on the left panel, while $h_z$ are varied and $D_z=0$ on the right panel.}
\label{fig:Energy_JD}
\end{center}
\end{figure}
In fact, this is easily and rigorously seen from the perfectly dimerized limit $\delta=1$. 
In this limit, the Hamiltonian with $d_x=0$ is reduced to the sum of independent two-site Hamiltonians, 
\begin{eqnarray} \label{eq:2body}
H_\textrm{two-site} = 2\vec{S}_1 \cdot \vec{S}_2 + D_z[(S^z_1)^2+(S^z_2)^2] + h_z(S^z_1-S^z_2). 
\end{eqnarray}
Again we can obtain the two trivial states $\left| D \right>$ and $\left| N \right>$ in the limits $D_z \rightarrow \infty$ and $h_z \rightarrow \infty$, respectively. 
Therefore, the continuity between $\left| D \right>$ and $\left| N \right>$ is confirmed by finding a path on which no level crossing occurs in the lowest energy spectrum between these limits. 
In Fig.~\ref{fig:Energy_JD}, we plot energy spectra of the two-site Hamiltonian. 
The singlet state at $D_z=h_z=0$ is adiabatically connected to both the states $\left| 00 \right>$ and $\left| -+ \right>$. 
From our numerical data we thus conclude that, by breaking the site-centered inversion symmetry, the two states $\left| D \right>$ and $\left| N \right>$ are adiabatically connected and thus no longer distinguished.

\end{document}